\documentclass[12pt]{iopart}

\usepackage{iopams}
\usepackage{graphicx}
\usepackage{appendix}
\begin{document}

\title{Optimisation of two-dimensional ion trap arrays for quantum simulation}

\author{James D. Siverns, Seb Weidt, Kim Lake, Bjoern Lekitsch, Marcus D. Hughes, and Winfried K. Hensinger}

\address{Department of Physics and Astronomy, University of Sussex, Brighton, UK\\
BN1 9QH}
\ead{W.K.Hensinger@sussex.ac.uk}
\begin{abstract}
The optimisation of two-dimensional (2D) lattice ion trap geometries for trapped ion quantum simulation is investigated. The geometry is optimised for the highest ratio of ion-ion interaction rate to decoherence rate. To calculate the electric field of such array geometries a numerical simulation based on a ``Biot-Savart like law" method is used. In this article we will focus on square, hexagonal and centre rectangular lattices for optimisation. A method for maximising the homogeneity of trapping site properties over an array is presented for arrays of a range of sizes. We show how both the polygon radii and separations scale to optimise the ratio between the interaction and decoherence rate. The optimal polygon radius and separation for a 2D lattice is found to be a function of the ratio between rf voltage and drive frequency applied to the array. We then provide a case study for $^{171}$Yb$^+$ ions to show how a two-dimensional quantum simulator array could be designed.
\end{abstract}


\section{Introduction}
Trapped ions possess long lived addressable internal states and can be highly decoupled from their environment. This makes them an important tool in the development of quantum information processing \cite{Cirac,Kielpinski} and quantum simulation \cite{Pons,Johanning,Gerritsma,Schneider1,Kim2}. When used for quantum simulation they enable complex spin systems, among others, to be investigated beyond the practical limitations of classical computation. For example, trapped ions have been used for quantum simulations of the evolution of paramagnetic into (anti-)ferromagnetic order in a spin system, \cite{Friedenauer,Islam} and frustrated anti-ferromagnetic Ising interactions, \cite{Kim,Korenblit}. These first simulations were carried out using one-dimensional trapping arrays and state dependant forces applied using laser beams.

More complex simulations will require ions trapped in 2D arrays and interaction schemes compatible with these. Advances towards 2D trapping arrays suitable for quantum simulations have been made by trapping ions in a millimetre-scale mechanically fabricated metal mesh \cite{Clark} and by the successful implementation of microfabrication techniques for ion traps \cite{Hughes1}. In addition, interaction schemes based on large oscillating or static magnetic field gradients have been proposed \cite{Chiaverini, Mintert1} and demonstrated in an on-chip microwave gate \cite{Ospelkaus,Khromova}. With the recent advances in the field, ion trap quantum simulations using large scale 2D ion trap lattices are within the reach of current technology.

In order to create a 2D array of trapped ions a repeating 2D surface geometry is required. Decoherance due to anomalous heating is a major issue for large scale quantum simulations. As this heating scales approximately as $r^{-4}$ \cite{Turchette}, where $r$ is the ion height above the trapping surface, it is advantageous for ions to be trapped high above the surface. However, when individual surface microtraps are placed together so that their separation is less than around twice the ion height the individual electric fields start to overlap and distort the resulting trapping fields \cite{Schmied1}. In extreme cases this can lead to the traps combining to produce a singular trapping zone. To compensate for this the electrode structure has to be altered when operating within this regime \cite{Schmied1}. Schmied \etal \cite{Schmied1} have investigated surface-electrode geometries and developed an algorithm that optimises geometries to maximise the electric field curvatures of individual trapping sites for arbitrary ion heights and separations. Individual trapping sites shown in \cite{Schmied1, Schmied2} were optimised using this algorithm leading to non-intuitive electrode patterns which can contain many isolated radio-frequency (rf) and static voltage electrodes. Another proposal \cite{Kumph1} working outside this regime uses rf electrodes with controllable rf voltages to lower trap frequencies and decrease ion-ion distances and, therefore, increase interaction strengths. However, this requires the use of multiple independent rf electrodes and individually controllable rf voltages posing an additional experimental complication.

In this paper we present an optimisation process for ion trap topologies based on a single island of rf electrode, reducing the requirement for buried rf wires and multiple rf electrodes. We focus on the development of an optimum lattice geometry where the ratio of coupling rate to the decoherence rate due to ion heating is maximised and made homogeneous across the lattice. This is achieved by minimising the secular frequencies of the trapping sites whilst, simultaneously, keeping the trapping depths above a minimum trap depth (for 
illustrative purposes we use 0.1eV) to allow for successful operation of the proposed 2D lattice designs. We will concentrate on the optimum lattice topology for hexagonal, square and centred rectangle lattices. An investigation is also carried out on how optimal geometries depend on the overall lattice size, and we discuss the choice of and scaling for experimental parameters such as rf voltage, drive frequency, ion mass and electric field noise density. 

Additionally it is possible that two-dimensional ion arrays of this type could also be used for quantum computation. For example, cluster state quantum computation could be carried out in such a system \cite{WunderlichH,Xu}. However, additional constraints may also have to be taken into account.

\section{Ion-ion interactions and Lattice Geometry}
\subsection{Ion-ion interactions}\label{sec:phys}
\begin{figure}[!htp]
\centering
\includegraphics[width=0.75\columnwidth]{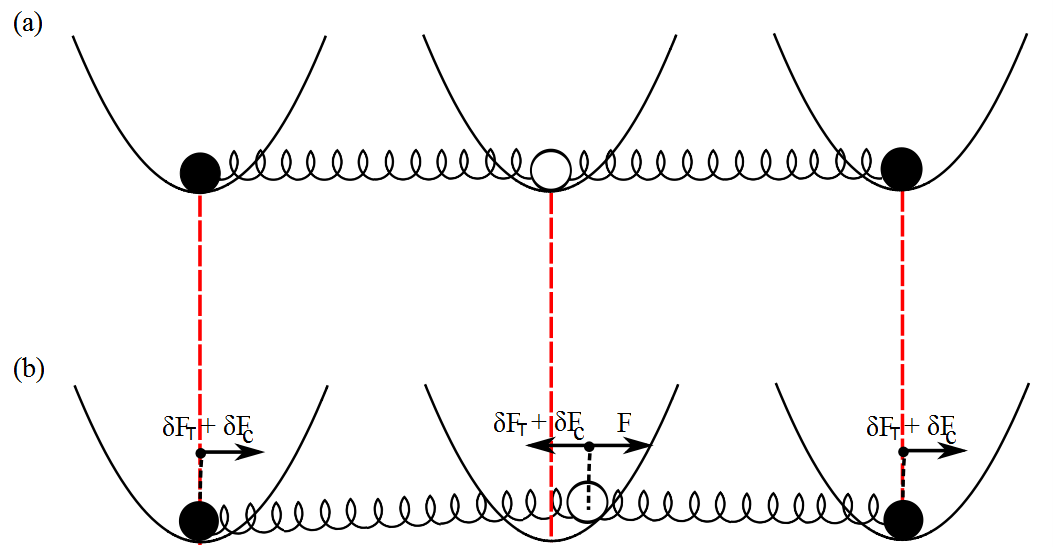}
\caption{Pictorial diagram of three ions in adjacent traps. The ions feel a Coulomb force indicated by the springs between each ion and can be placed in two different states indicated by their colour. (a) Pictorial diagram showing the case with no state dependent force present. (b) Pictorial diagram showing how the system reacts to the presence of a state dependent force, $F$. Here the ions feel a change in the Coulomb force, $\delta F_{C}$, due to the displacement of the ions and a change in the restoring force, $\delta F_{T}$.}
\label{new}
\end{figure}

Two-dimensional lattices of ions can be used as a quantum simulator for many body spin-$1/2$ systems \cite{Porras1}. Forces such as the trapping potential, $F_{T}=-m\omega^{2}_{i}$, and the coulomb force, $F_{C}=-e^{2}/(4\pi\epsilon_{0}A^{2})$, between the individual ions determine the equilibrium position. Laser beams or magnetic field gradients can be used to impart an additional force to the ions which displace the ion(s) depending on its internal state. This displacement leads to a change in the coulomb force and thereby displaces the neighbouring ion(s) dependent on their own internal state. This state dependant coupling is given by \cite{Porras1}

\begin{equation}\label{eq:1}
J = \frac{\beta F^{2}}{m\omega^{2}}
\end{equation}

\noindent
where $m$ is the mass of an ion, $F$ is the magnitude of the state dependant force applied to each ion and $\omega$ is the trap's secular frequency. We will consider how this force is applied later in the article. $\beta$ is the ratio of the change in the Coulomb force to the change in the restoring force due to the displacement of the ions caused by the state dependant force and is given by \cite{Porras1}

\begin{equation}\label{eq:2}
\beta = \frac{e^{2}}{2\pi\epsilon_{0}m\omega^{2}A^{3}},
\end{equation}

\noindent
where $A$ is the ion-ion spacing. There are two cases to consider, when $\beta>1$ the change in the Coulomb force, $\delta F_{C}$, due to the displacement of an ion is dominant over the change in the restoring force, $\delta F_{T}$. This results in an interaction over a large number of trapping sites. When $\beta<1$, the opposite is true resulting in an interaction which decays rapidly across the array. Trapping ions in a 2D array of microtraps makes it possible to satisfy the condition that $\beta <1$ allowing systems with short range interactions to be simulated. An illustration of this system is shown in figure \ref{new}.

It is important to consider sources of decoherence when designing a 2D ion trap array. The internal state of an ion can remain coherent for 10's of seconds \cite{Langer1, Haffner1}. However, motional decoherence due to anomalous heating of ions will be an important factor during quantum operations within small scale ion traps as the implementation of spin dependant couplings involves the use of motional states of the ion. The coupling, $J$, will be observable if the coupling time, $T_{J}=1/J$, is less than the motional decoherence time in the system and, therefore, the ratio of these two times is an important parameter of the system and is given by

\begin{equation}\label{ksim}
K_{sim}=\frac{T_{\dot{n}}}{T_{J}}
\end{equation}

\noindent
where \cite{Turchette},

\begin{equation}
T_{\dot{n}}=\frac{4m\omega\hbar}{e^{2}S_{E}(\omega)}.
\end{equation}

\noindent
Here $S_{E}(\omega)$ is the electric field noise density \cite{Turchette, Hughes1}. In order for an interaction to occur on faster time-scales than the decoherence in the system, we require $K_{sim}>1$ and it is the aim of the optimisation process presented in this work to optimise the geometry in order to maximise this parameter. To acquire an understanding of how a geometry can affect K$_{sim}$ it is necessary to determine its form with respect to the geometry variables. The form of $T_J$ can be found by substituting equation \ref{eq:2} into \ref{eq:1} and is given by

\begin{equation}\label{interact_time1}
T_J=\frac{2\pi\epsilon_{0}m^{2}\omega^{4}A^{3}\hbar}{e^{2}F^2}.
\end{equation}

\noindent
The K$_{sim}$ parameter can then be expressed as

\begin{equation}\label{ksim_mass_sec1}
K_{sim}=\frac{2F^2}{S_{E}(\omega)\pi\epsilon_{0}m\omega^{3}A^{3}}.
\end{equation}

\noindent
The secular frequency, $\omega$, of a trapped ion \cite{Madsen} can be expressed as a function of $\alpha$ defined as

\begin{equation}\label{alpha_def}
\alpha=\frac{V}{\Omega}
\end{equation}

\noindent
where $V$ is the amplitude of the rf voltage applied to the trap and $\Omega$ is 2$\pi$ times the drive frequency in Hz, yielding,

\begin{equation}\label{secfr_mass_sec}
\omega=\frac{eV\eta_{geo}}{\sqrt{2}m\Omega r^2}=\frac{e\alpha\eta_{geo}}{\sqrt{2}m r^2}
\end{equation}

\noindent
where $r$ is the height of an ion above the surface, $e$ is the charge of an electron and $\eta_{geo}$ is an efficiency factor which can range between zero and one depending on the form of the geometry \cite{Madsen}.

The secular frequency given in equation \ref{secfr_mass_sec} can then be used along with $S_{E}(\omega)=\Xi r^{-4}\omega^{-1}$, where $\Xi$ is a coefficient that can be experimentally obtained and depends on the temperature and surface of the trap electrodes (see \cite{Hughes1} for a listing) to re-express equation \ref{ksim_mass_sec1} in the form

\begin{equation}\label{ksim_alpha_sub}
K_{sim}=\frac{4F^2mr^8}{\Xi\alpha^2\eta_{geo}^2\pi\epsilon_0A^3}.
\end{equation}

\noindent
To further understand how the geometry effects the K$_{sim}$ value we will now introduce the parameters of a lattice geometry and relate them to equation \ref{ksim_alpha_sub}.
\subsection{Two-dimensional ion trap lattice geometry}\label{sec:geo}
\begin{figure}[!htp]
\centering
\includegraphics[width=0.75\columnwidth]{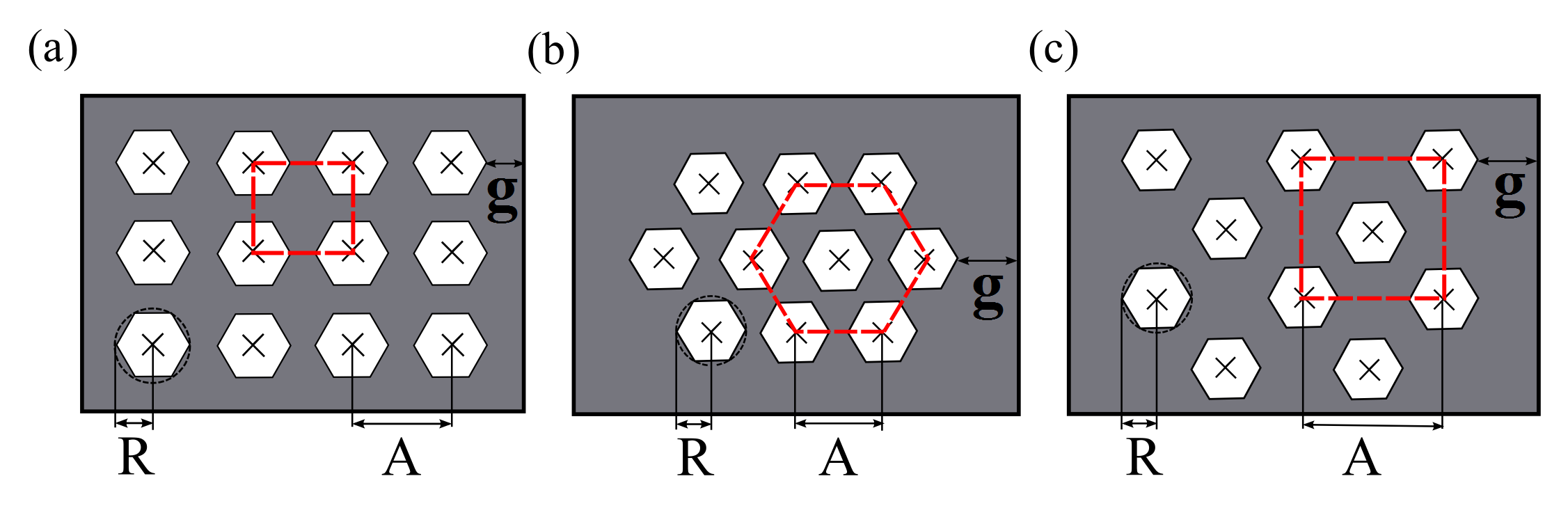}
\caption{Diagrams showing the polygon radii $R$, the separation between the polygon centrers, $A$, and the distance between the last polygon and the edge of the rf electrode (shown in grey), $g$. (a) Diagram showing a three by four ion trap surface array consisting of six sided polygons arranged with square unit cells. (b) Diagram showing a similar surface array arranged into hexagonal unit cells. (c) Diagram showing a surface array arranged into centred rectangular unit cells. The unit cells are indicated by dashed lines.}
\label{array}
\end{figure}

A lattice is a regular tiling of a space by a primitive unit cell. Previous works \cite{Clark, Kumph1} concentrate on lattices created from square unit cells. In total there exist five types of cell which can be used to form a 2D lattice: centred rectangular, hexagonal and square as shown in figure \ref{array}, and rectangular and oblique \cite{Kittel1}. The rectangular and oblique structures are not considered in this work due to their non-uniform ion-ion distances.

Figure \ref{array} shows the polygon-polygon separation which is equal to the ion-ion distance, $A$, in equation \ref{ksim_alpha_sub}. The polygon radius, $R$, along with the separation, will determine the height above the surface at which the ion is trapped, $r$, with larger polygon radii yielding higher ion heights. Another variable to be considered is the gap between the outer polygon in the array to the edge of the rf electrode, $g$. This can be used to alter the homogeneity of the individual trapping sites within the array. In general a non-homogeneous system results in spin dependant coupling rates which are a function of the lattice site, posing a significant problem for the scalability of such an array \cite{McHugh1}.

\section{Simulation of lattices}\label{sec:error}
To determine the electric field produced by a two-dimensional array a method based on the Biot-Savart like law described by Oliveira and Miranda \cite{Mario1} was used. This method calculates the electric field produced by an arbitrarily shaped two-dimensional electrode which is held at a potential V whilst the rest of the plane is held at a potential of zero. The electric field observed at a given point, $\overline{X}$, in space due to such an area held at a potential and bounded by a path $C$ is given by \cite{Mario1}

\begin{equation}\label{bs_electric_field}
E_{(\overline{X})}=\frac{V}{2\pi}\oint_{C}\frac{(\overline{x}-\overline{x}')\times d\overline{s}}{\left|\overline{x}-\overline{x}'\right|^{3}}
\end{equation}

\noindent
where the curve, $C$, bounds the electrode and $\overline{x}'$ and $\overline{x}$ are vectors that locate the source point and field point respectively. By calculating the electric field in this manner an assumption is made that there is no gap between the areas held at the potential, $V$, and the areas held at zero. In microfabricated surface traps, gaps between the electrodes are required and typically range from 3$\mu$m - 10$\mu$m \cite{Hughes1}. If, however, these gaps are small in comparison to the electrode structures they will not alter the trapping fields significantly \cite{Schmied2, House1, Amini1}. The electric fields of individual electrodes can then be combined to determine potential nils and, therefore, trapping positions in the 2D trap arrays, using the numerical Gauss-Newton algorithm. The secular frequencies, trap depths and ion heights at these positions can then be determined.

To calculate the error of our numerical integration we compared simulations of five wire symmetric surface trap geometries with different central static electrode and rf rail widths in the gapless plane approximation using the method of Oliveira and Miranda \cite{Mario1} to results obtained with analytical equations described by House \cite{House1}. In all the geometries simulated the two rf rails were of equal width. Similarly to House, the outer static voltage electrodes were approximated as an infinitely long ground plane although the length of the inner rails were set to 3000 $\mu$m instead of infinite. A selection of these simulation results are shown in table \ref{house_compare}.

\begin{table}[ht]
\centering
  \begin{tabular}{| c | c | c | c || c | c || c | c | }
    \hline
     \multicolumn{4}{|c||}{Electrode parameters}  & \multicolumn{2}{|c||}{Simulations} & \multicolumn{2}{|c|}{House equations} \\ \hline
    rf width & central static  & rf Volt.  & rf freq.  & r  & $\omega$  & r  & $\omega$     \\
     $[\mu$m$]$ & electrode width & $[$V$]$ & $[$MHz$]$ & $[\mu$m$]$ & $[$MHz$]$ & $[\mu$m$]$ & $[$MHz$]$    \\
     & $[\mu$m$]$ & &  &  &  &  &      \\ \hline
    100  & 50   & 250  & 75  & 55.8   & 6.86  & 55.9  & 6.87                         \\ \hline
    100  & 50   & 500  & 60  & 55.8   & 4.29  & 55.9  & 4.29                         \\ \hline
    200  & 100  & 250  & 30  & 110.1  & 2.17  & 111.8 & 2.20                         \\ \hline
    200  & 100  & 500  & 40  & 110.1  & 3.26  & 111.8 & 3.30                         \\ \hline
    500  & 150  & 250  & 20  & 165.4  & 1.48  & 167.7 & 1.52                         \\ \hline
    500  & 150  & 500  & 25  & 165.4  & 2.37  & 167.7 & 2.43                         \\ \hline
   \end{tabular}
   \caption{Table showing the secular frequency, $\omega$, and ion height, $r$, for different five wire surface trap geometries as calculated by the analytical method in House \cite{House1} and simulated by the method used in this work based on the Biot-Savart like law \cite{Mario1}.}
   \label{house_compare}
\end{table}

In these results a general error for the ion heights and secular frequencies of less than 2$\%$ and 3$\%$ respectively was found, which leads to a maximum error in $K_{sim}$ of 10$\%$. For the following simulations it is therefore assumed that the maximum $K_{sim}$ error is 10$\%$. Additionally, numerical simulations of the geometries were carried out using methods described in \cite{Singer}, which indicate similar errors and trends for the ion height and secular frequency as the results obtained with the Biot-Savart like method.

\section{Lattice geometry optimisation}
In this section we show how the parameters of the lattice geometry (discussed in section \ref{sec:geo}) can be optimised to achieve the highest possible K$_{sim}$ value across the array for a given set of experimental parameters. To do this, we first show how to maximise the homogeneity of individual site properties over an array by varying the distance between the outer polygon in the array to the edge of the rf electrode, $g$, and show how this scales with lattice size. We will then use the homogeneous arrays to calculate the optimum number of sides, $n$, a polygon within the array should possess in order to maximise K$_{sim}$. We then outline a method for the optimisation of the polygon radii, $R$, and separation, $A$, of an array and show how these vary with increased lattice size and ion mass.

\subsection{Increasing the homogeneity of K$_{sim}$ across the array}\label{sec:optimise_g}
\begin{figure}[!htp]
\centering
\includegraphics[width=0.75\columnwidth]{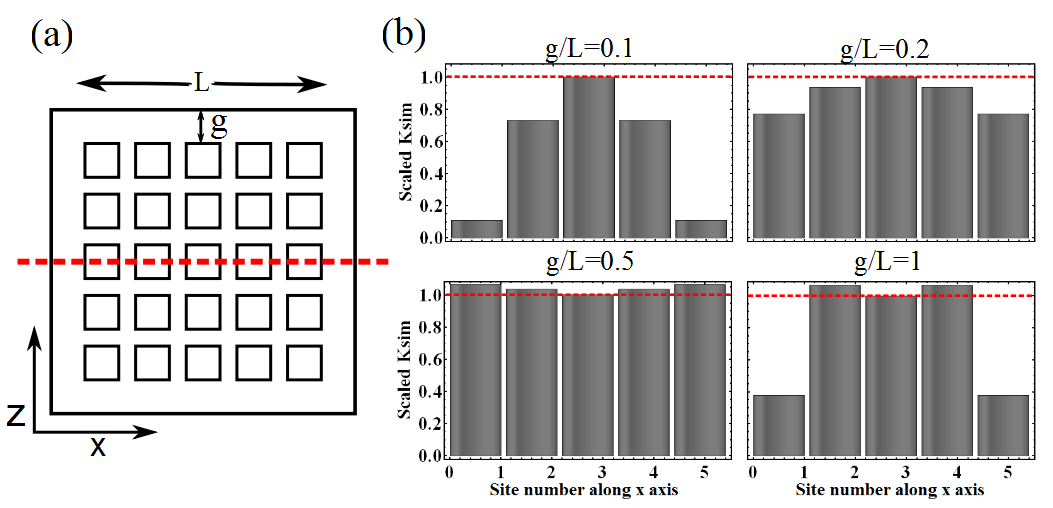}
\caption{Diagram showing the effect of varying the distance g on the scaled K$_{sim}$ value of the individual trapping sites. The K$_{sim}$ values shown are scaled with that of the central site. (a) Representation of a 5 by 5 square type lattice array indicating the axis labelling. (b) Slices across the array (indicated by the dotted line in (a)) for g/L values of $0.1$, $0.2$, $0.5$ and $1$.}
\label{bowl_flatten}
\end{figure}

This is achieved by ensuring homogeneous secular frequencies, ion heights and trap depths across all the array sites. As shown in figure \ref{bowl_flatten} the K$_{sim}$ of trapping sites in an array can be altered to approach a common value if the distance, $g$, between the edge of the outer polygon and the edge of the rf electrode containing the polygon array is adjusted. As the value of $g$ is increased, the $K_{sim}$ value of the sites towards the centre drop, however, the outer sites K$_{sim}$ value rises, resulting in the properties converging towards each other. If the distance g is increased further beyond the point at which maximum homogeneity occurs the outer site properties drift away from those of the central sites and, therefore, decrease the homogeneity of the array.

To provide a value of g which is universal for all lattice sites, its value is given in units of lattice side length, $L$.  The lattice side length is determined by the polygon separation, $A$, and radius, $R$, and can be expressed as

\begin{equation}\label{lattice_side}
L=(M-1)A+2R,
\end{equation}

\noindent
where $M$ is the number of lattice sites along one side of an array.

\noindent
In order to quantify the arrays homogeneity, $H$ is defined as the average deviation of K$_{sim}$ of each lattice site from the K$_{sim}$ of the central site and is given by

\begin{equation}\label{H}
H = \frac{1}{N} \sum_{n=1}^{N} \left|1-\frac{K_{sim_{n}}}{K_{sim_{centre}}}\right|
\end{equation}

\noindent
where $N$ is the total number of trapping sites in the lattice.

\begin{figure}[!htp]
\centering
\includegraphics[width=0.75\columnwidth]{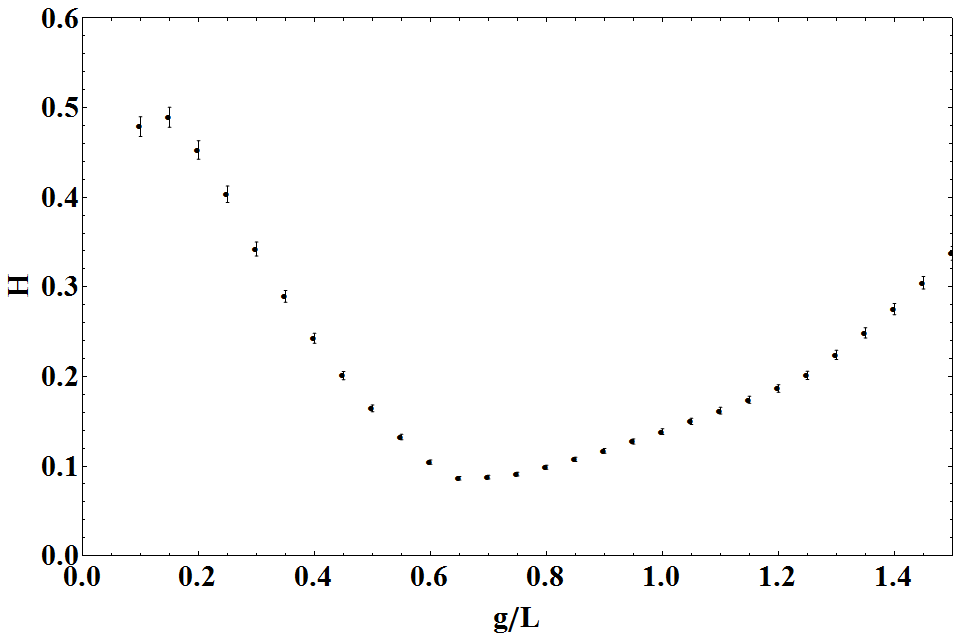}
\caption{Graph showing the average deviation of the K$_{sim}$ of each lattice site from the K$_{sim}$ of the central site, $H$, for a five by five square type unit cell lattice for $0<g/L<1.5$. The error on $H$ is given by $\frac{0.13}{\sqrt{N}}H$ and the error of the minimum of $H$ is determined by observing the spread of g/L which agrees, within error, with the minimum position.}
\label{gL_example}
\end{figure}

Figure \ref{gL_example} shows $H$ for a five by five square type unit cell lattice for $0<g/L<1.5$. The maximum homogeneity, and thus the optimum g/L, is found when $H$ is minimised. The error associated with $H$ is given by

\begin{equation}\label{error_H}
\sigma_{H} = \frac{\sqrt{N(\sigma_{Ksim})^{2}+N(\sigma_{Ksim_{centre}})^{2}}}{N}
\end{equation}

\noindent
where the error on all K$_{sim}$ values is 10$\%$, as shown in section \ref{sec:error}. This yields an overall percentage error on $H$ of 0.13/$\sqrt{N}\%$.

\begin{figure}[!htp]
\centering
\includegraphics[width=0.75\columnwidth]{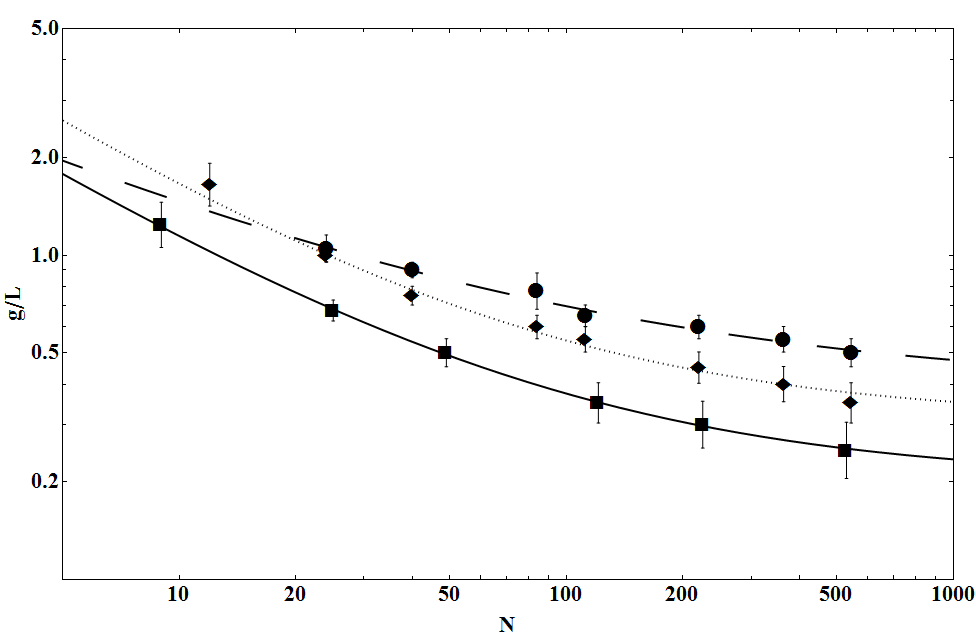}
\caption{Graph showing the optimum g/L as a function of the total number of trapping sites, N, for square lattices (square markers), hexagonal lattices (circular markers) and centre rectangular lattices (diamond markers). The curves are given by $g/L=a+bN^{-B}$ with $a$, $b$ and $B$ values for different lattice types shown in table \ref{g/L_scale}.}
\label{optimumg}
\end{figure}

Figure \ref{optimumg} shows the optimum g/L for hexagonal, central rectangular and square unit cell lattices of different sizes. The curves are found to be described by an equation of the form $g/L=a+bN^{-B}$ with $a$, $b$ and $B$ values for different lattice types shown in table \ref{g/L_scale}. For large lattices, $g/L$ is independent of $N$, as  trapping sites close to the edge of the lattice are influenced only by the electric field created by that edge. In small lattices, however, the optimum g/L increases as the effect of the electric field from the opposite edge of the lattice increases.

\begin{table}
\begin{center}
    \begin{tabular}{ | l || c | c | c |}
    \hline
    Lattice Type  & a & b & B  \\ \hline
    Square & 0.20$\pm$0.01 & 5.21$\pm$0.38 & 0.74$\pm$0.03 \\ \hline
    Hexagonal & 0.39$\pm$0.08  & 3.76$\pm$1.40 & 0.54$\pm$0.14 \\ \hline
    Centre Rectangular & 0.31$\pm$0.04 & 7.84$\pm$2.43 & 0.76$\pm$0.11 \\ \hline
    \end{tabular}
\end{center}
\caption{Table showing $a$ and $b$ values for the fits which describe $g/L$ as a function of the number of sites in the lattice.}
\label{g/L_scale}
\end{table}

\subsection{Optimising the number of polygon sides} \label{sec:optimise_n}
We now investigate the optimum number of polygon sides, $n$, providing the highest $K_{sim}$ value on the central trapping site (located above the central polygon) of a lattice.

To ensure the results are universal for all lattice geometries, g/L is set to the value which maximises the homogeneity of each array, and all other parameters are scaled by normalising K$_{sim}$ to that of an identical geometry with polygons of 100 sides. This also allows comparison between the different types of lattices.

\begin{figure}[!htp]
\centering
\includegraphics[width=0.75\columnwidth]{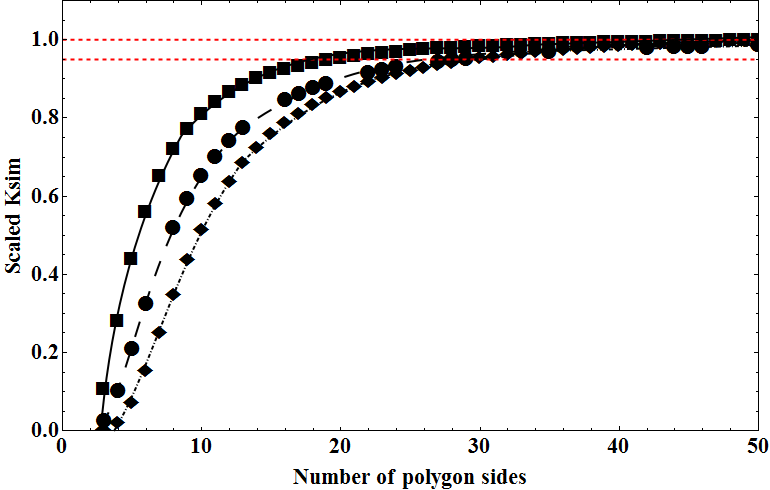}
\caption{Graph showing the relationship between the number of polygon sides and K$_{sim}$ for square (square markers), hexagonal (circular markers) and central rectangular (diamond markers) unit cell lattices. The dashed lines show the asymptotes of 1 and 0.95 of the scaled K$_{sim}$ value.}

\label{n_vary}
\end{figure}

Figure \ref{n_vary} shows the scaled K$_{sim}$ for the central site as the number of polygon sides is varied. As the number of sides is increased the value of K$_{sim}$ approaches an asymptote, shown by the upper dashed line. This indicates that the best geometry will be made from circular electrodes. However, simulation times grow with increasing polygon side number and so it is advantageous to reduce this number to a minimum. It is shown that $\approx$95$\%$ of the maximum achievable K$_{sim}$ (indicated by the lower dashed line) can be achieved with around $\geq$25-30 sides in the polygons.

\subsection{Optimisation method for polygon separation and radius}\label{sec:optimise_meth}
In this section we now maximise the K$_{sim}$ of any arbitrary geometry. We will then go on and determine optimum geometries and show how they scale and, ultimately, are determined by $\alpha=V/\Omega$. We use the value of g determined in section \ref{sec:optimise_g} in order to provide the maximum K$_{sim}$ homogeneity across the lattice, and set the number of sides to 25 as this provides a good approximation to the optimum circular geometry while keeping the simulation time at a minimum, as shown in section \ref{sec:optimise_n}.

When considering any fixed arbitrary geometry, equation \ref{ksim_alpha_sub} shows that K$_{sim}$ can be maximised by reducing the value of $\alpha$. However, the minimum achievable $\alpha$ is limited by the lowest usable trap depth, as the trap depth is proportional to $\alpha^2$  and is given by

\begin{equation}\label{house_depth}
T_{D}=\frac{\zeta e^2\alpha^2}{\pi^2m}
\end{equation}

\noindent
where $e$ is the charge of an ion and $\zeta$ is a geometrical factor which is a function of $A$ and $R$ \cite{House1}.

\begin{figure}[!htp]
\centering
\includegraphics[width=0.75\columnwidth]{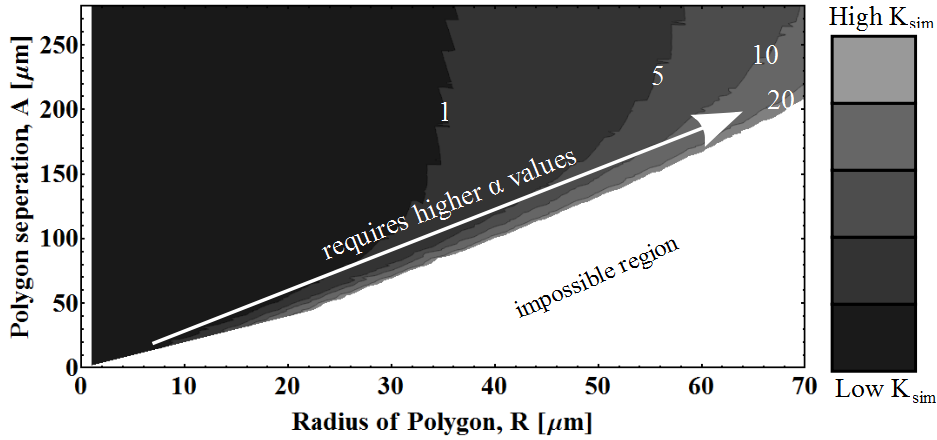}
\caption{Example graph showing how the K$_{sim}$ (absolute values indicated by numbers on contour lines) varies as a function of polygon radius and separation. The graph also indicates that the value of $\alpha$ increases as the radius and separation are increased. This data was obtained using the method described in section \ref{sec:optimise_meth} with a polygon separation and radius resolution of 1$\mu$m and a minimum trap depth of 0.1 eV. The value of $\alpha$ in the figure ranges from zero to $\approx$1.5 VMHz$^{-1}$. The impossible region describes geometries where individual trapping sites start to combine to a single one and so posses a polygon radius, $R$, greater than or equal to a third of the polygon separation, $A$.}
\label{A_R_Ksim_plot}
\end{figure}

We will now focus on finding optimal geometries which we define as geometries, which yield the highest value of K$_{sim}$ for a given value of $\alpha$. This will be carried out by fixing the trap depth at a reasonable minimum value (we use 0.1 eV for illustration purposes) which, as discussed above, provides the maximum K$_{sim}$ for a given geometry, and then investigating the dependence of polygon radius, separation and ion height with $\alpha$. To determine these dependencies a K$_{sim}$ contour plot is made by calculating the K$_{sim}$ over a range of polygon separation, $A$, and radius, $R$, with a resolution of 1$\mu m$. The range of polygon radius and separation used should not create traps with inter-well barriers of less than the minimum trap depth value, and to ensure this the polygon radius was kept to less than a third of the polygon separation.  For each combination of polygon separation and radius a value of $\alpha$ is found which yields the minimum trap depth and, thus, maximises the K$_{sim}$ of the particular geometry. By following this method one can obtain the $\alpha$ required to achieve the minimum trap depth, the ion height, $r$, and K$_{sim}$ for each geometry. From the resulting data the polygon separation and radius which yields the highest K$_{sim}$ for a given $\alpha$ (the optimum geometry) can then be found. A graphical example of such data is shown in figure \ref{A_R_Ksim_plot}.

It can be seen from this method and the example data in figure \ref{A_R_Ksim_plot} that the highest K$_{sim}$ will be achieved with an infinite value of $\alpha$, $R$ and $A$. However, other effects may limit the magnitude of $\alpha$. In order to determine a limit on $\alpha$ it is, therefore, necessary to describe the various array and trapping field dependant properties (such as secular frequency, ion height and K$_{sim}$) in terms of $\alpha$.

\begin{figure}[!htp]
\centering
\includegraphics[width=1\columnwidth]{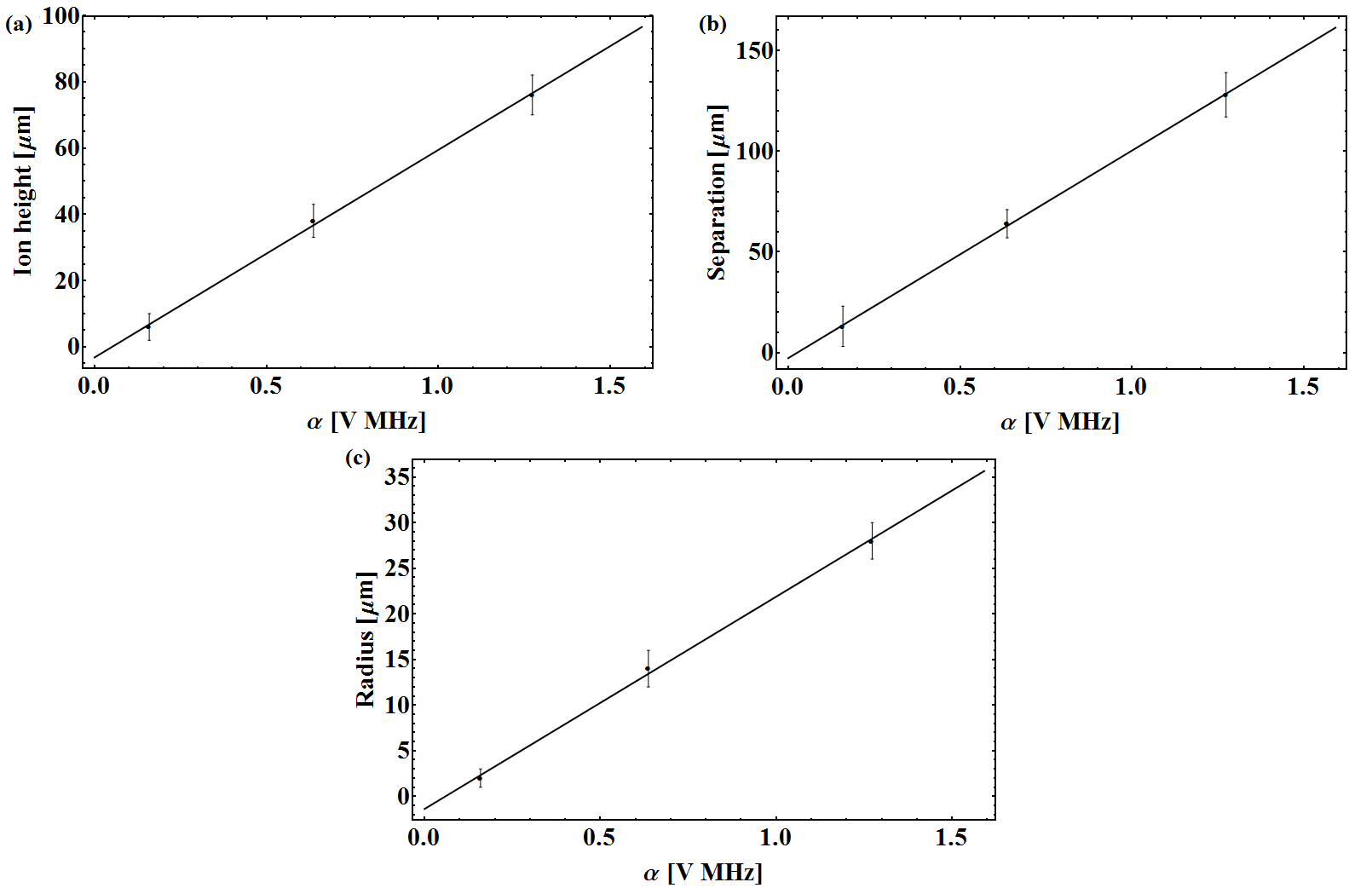}
\caption{Graphs showing the ion height (a), polygon separation (b) and polygon radius (c) of an optimised lattice as a function of the ratio $\alpha$. In all cases the plots are shown using $\alpha=V/\Omega$ where $\Omega$ is 2$\pi$ times the drive frequency in Hz, and for arrays made from square type unit cells of polygons with 81 sites and for $^{171}$Yb$^+$ ions.}
\label{thingsValpha}
\end{figure}

\begin{figure}[!htp]
\centering
\includegraphics[width=1\columnwidth]{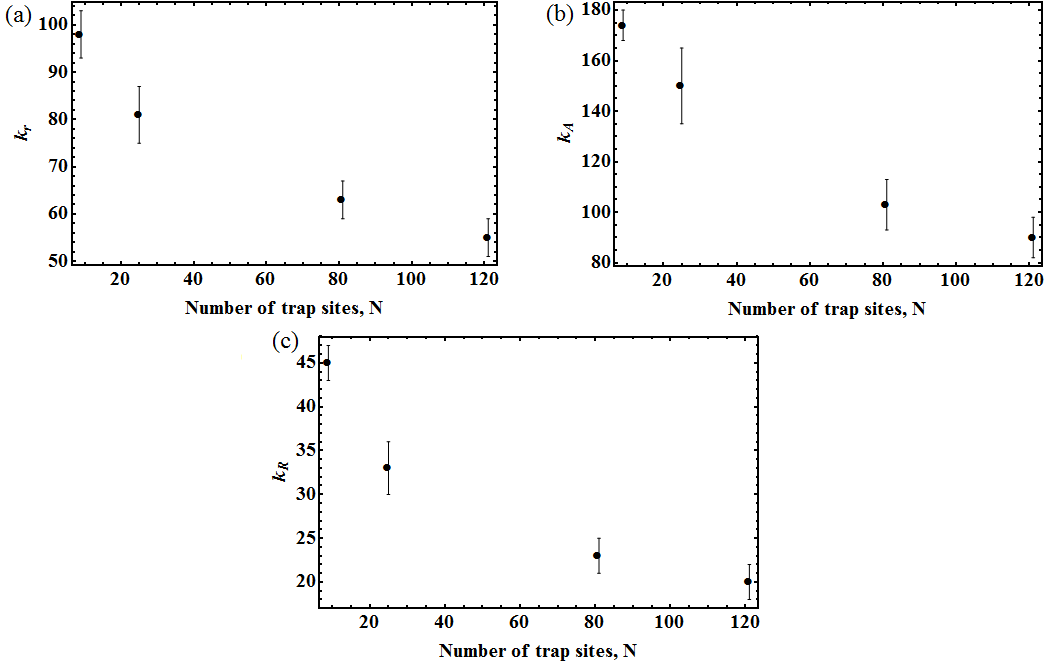}
\caption{Graphs showing the value of $k_r$ (a), $k_A$ (b) and $k_R$ (c) as a function of the number of trapping sites, $N$. In all cases the plots are shown using $\alpha=V/\Omega$ where $\Omega$ is 2$\pi$ times the drive frequency in Hz and are for arrays made from square type unit cells of polygons using $^{171}$Yb$^+$ ions.}
\label{ksVN}
\end{figure}

By plotting these optimum parameters (polygon radius, separation and ion height) as a function of $\alpha$, as shown in figures \ref{thingsValpha}(a), (b) and (c), linear relationships of the form

\begin{equation}\label{r_alpha}
r=k_r\alpha
\end{equation}

\begin{equation}\label{A_alpha}
A=k_{A}\alpha
\end{equation}

\noindent
and

\begin{equation}\label{R_alpha}
R=k_{R}\alpha
\end{equation}

\noindent
are found for the optimal geometries. The values of $k_r$, $k_A$ and $k_R$ are dependant on the number of trapping sites in an array, as shown in figures \ref{ksVN}(a),(b) and (c) respectively, for lattices made from square type unit cells of polygons.

It is important to stress that equations \ref{r_alpha}, \ref{A_alpha} and \ref{R_alpha} are only valid in the case of optimal geometries, which depend solely on $\alpha$. With this in mind, it is now possible to re-express the secular frequency in equation \ref{secfr_mass_sec} to describe the secular frequency of an optimised geometry as:

\begin{equation}\label{sec_freq_opt}
\omega=\frac{e\eta_{geo}}{\sqrt{2}k_{r}^{2}m\alpha}.
\end{equation}

K$_{sim}$ in equation \ref{ksim_alpha_sub} can also be re-expressed to describe that of an optimised geometry:

\begin{equation}\label{KsimValpha}
K_{sim}=\frac{4F^2mk_{r}^{8}\alpha^3}{\Xi\eta_{geo}^2\pi\epsilon_{0}k_{A}^{3}}.
\end{equation}

\noindent
Equation \ref{KsimValpha} shows that, for optimal geometries, K$_{sim}$ is proportional to $\alpha^3$ (as the value of $\alpha$ determines the electrode dimensions to be used) and so, to produce an array with a high K$_{sim}$ for a given number of lattice sites (as $k_r$ and $k_A$ are a function of the number of trapping sites), a large value of $\alpha$ is preferable. Equations \ref{A_alpha} and \ref{R_alpha} show that the optimum geometry size is proportional to $\alpha$. It, therefore, follows that larger lattice geometries will produce larger values of K$_{sim}$. This effect is illustrated in figure \ref{A_R_Ksim_plot} which shows the K$_{sim}$ as a function of polygon radius and separation. For optimised lattices, the optimal radius and separation will fall on a line described by $A=\left(k_A/k_R\right)R$, with higher values of $\alpha$ required for higher values of separation and radius as shown in equations \ref{A_alpha} and \ref{R_alpha} respectively.

The heating rate in ion traps has a strong dependency on the ion height ($\propto 1/r^{-4}$) \cite{Turchette}. A large K$_{sim}$ is achieved with large values of $\alpha$, resulting in large ion heights, as shown in equation \ref{r_alpha}. It, therefore, can be concluded that a different scaling of the heating rate (for example in cryogenic systems) does not change the optimisation process and optimal geometries.

It has now been shown that the optimal geometry for a given minimum trap depth and ion mass is determined solely by the value of $\alpha$.

Optimal geometries and their K$_{sim}$ values (in units of $1/\alpha^3$) can now be found by creating contour plots (such as shown in figure \ref{A_R_Ksim_plot}) for different values of lattice size, $N$, lattice unit cell type and ion mass, $m$. The error on the K$_{sim}$ value, calculated from equation \ref{ksim_alpha_sub}, was determined to be $\pm$10$\%$ by comparing the secular frequency and ion heights obtained using the program with those predicted by House's analytical solutions for a five wire surface trap geometry \cite{House1}.
\section{Optimisation results and analysis}
In this section, optimum polygon separations, A, and radii, R, are obtained using the method outlined above for square, hexagonal and centre rectangular unit cell type lattices. These are shown as function of lattice sizes and ion masses with the experimental constraint, $\alpha$, scaled out. Throughout this optimisation example, $^{171}$Yb$^+$ ions will be used unless otherwise stated.

\begin{figure}[!htp]
\centering
\includegraphics[width=1\columnwidth]{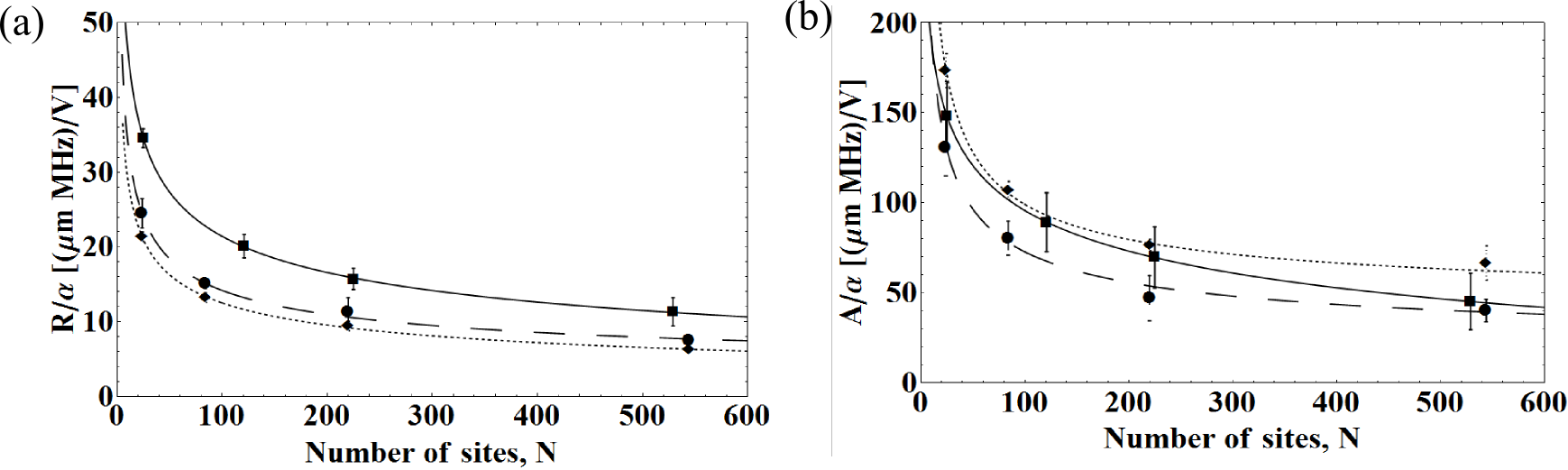}
\caption{(a) Graph showing how the optimum polygon radius, $R/\alpha$, varies as a function of the number of sites. (b) Graph showing how the optimum polygon separation, $A/\alpha$, varies as a function of the number of sites. For both (a) and (b) the results shown are for square (square markers), hexagonal (circular markers) and centre rectangular (diamond markers) unit cell lattices with $^{171}$Yb$^+$ ions.}
\label{opto_params}
\end{figure}

\begin{figure}[!htp]
\centering
\includegraphics[width=0.75\columnwidth]{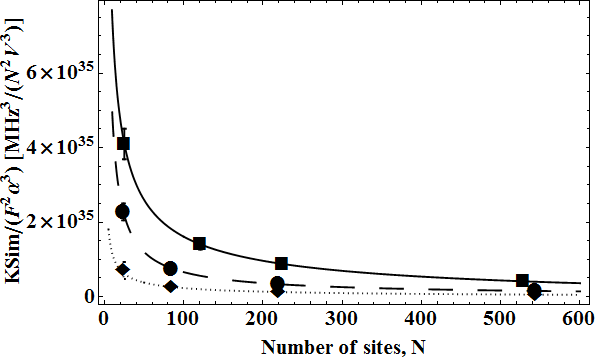}
\caption{Graph showing how the optimum K$_{sim}$/(F$^{2}\alpha^3)$ varies as a function of the number of sites for optimum lattices with $^{171}$Yb$^+$ ions. This is shown for square (square markers), hexagonal (circular markers) and centre rectangular (diamond markers) unit cell lattices. Here F is a state dependant force applied to the ions in the lattice.}
\label{opto_ksim}
\end{figure}

Figure \ref{opto_params} shows how the optimum scaled radius, $R/\alpha$, and separation, $A/\alpha$, of polygons vary as a function of lattice size for $^{171}$Yb$^+$ ions. As explained in the previous section we have assumed a minimum trap depth of 0.1 eV for illustrative purposes. It can be seen from this figure that as the size of the lattice increases, the optimum polygon radius and separation asymptotically tend towards values representative of an infinitely large lattice. This is expected as once a lattice becomes large enough, the addition of extra lattice sites will represent only a small change in the overall electrode geometry and, therefore, produce a small change in the electric field produced by the geometry. When the lattice is small however, additional lattice sites will represent a larger change in the geometry and will, therefore, cause a larger change in the electric field. Figure \ref{opto_ksim} shows how the scaled K$_{sim}$/($F^2\alpha^3$) scales as a function of the number of lattice sites, $N$, using scaled optimum polygon radii, $R/\alpha$, and separations, $A/\alpha$. The state dependant force $F$ will be considered in more detail in section \ref{quant_sim_error}. Due to the dependency of $K_{sim}$ on the geometry, the relationship between K$_{sim}$/(F$^{2}\alpha^3)$ and the number of sites is expected to be of similar form to that for optimal polygon radii, $R/\alpha$, and separation, $A/\alpha$, with the maximum K$_{sim}$/(F$^{2}\alpha^3)$ asymptotically tending towards a value representative of an infinitely large lattice.

Using the data shown in figures \ref{opto_params}(a), \ref{opto_params}(b) and \ref{opto_ksim} the optimum radii and separation of the polygons were found to follow a $c+dN^{-E}$ and $f+gN^{-G}$ relationship, respectively, while the maximum K$_{sim}$/(F$^{2}\alpha^3)$ follows a $k+lN^{-Q}$ trend. The values of $c$, $d$, $E$, $f$, $g$, $G$ $k$, $l$, and $Q$ are shown in tables \ref{table_Vn1}, \ref{table_Vn2} and \ref{table_Vn3}.

\begin{table}
\begin{center}
    \begin{tabular}{ | l || c | c | c | c |}
    \hline
    Lattice Type  & c & d & E\\ \hline
    Square & -5$\pm$1 & 101$\pm$7 & 0.29$\pm$0.03\\ \hline
    Hexagonal & 1$\pm$3 & 85$\pm$6 & 0.40$\pm$0.08\\ \hline
    Centre Rectangular & -2$\pm$2 & 68$\pm$6 & 0.34$\pm$0.05\\ \hline
    \end{tabular}
\end{center}
\caption{Table showing $c$, $d$ and $E$ values for the fits which describe $R/\alpha$ as a function of the number of sites in the lattice.}
\label{table_Vn1}
\end{table}

\begin{table}
\begin{center}
    \begin{tabular}{ | l || c | c | c | c |}
    \hline
    Lattice Type  & f & g & G \\ \hline
    Square & -136$\pm$7 & 457$\pm$5 & 0.15$\pm$0.01\\ \hline
    Hexagonal & 13$\pm$24 & 547$\pm$288 & 0.48$\pm$0.21\\ \hline
    Centre Rectangular & 40$\pm$15 & 831$\pm$282 & 0.57$\pm$0.13\\ \hline
    \end{tabular}
\end{center}
\caption{Table showing $f$ and $g$ and $G$ values for the fits which describe $A/\alpha$ as a function of the number of sites in the lattice.}
\label{table_Vn2}
\end{table}

\begin{table}
\begin{center}
    \begin{tabular}{ | l || c | c | c | c |}
    \hline
    Lattice Type  & k & l & Q \\ \hline
    Square & -(2.69$\pm$2.97)$\times$10$^{34}$ &  (3.09$\pm$1.31)$\times$10$^{36}$ & (0.61$\pm$0.13) \\ \hline
    Hexagonal & (0.52$\pm$6.00)$\times$10$^{33}$ &  (3.51$\pm$1.54)$\times$10$^{36}$ & (0.86$\pm$0.12) \\ \hline
    Centre Rectangular &  -(3.09$\pm$6.12)$\times$10$^{35}$ & (5.23$\pm$5.11)$\times$10$^{37}$ & (0.65$\pm$0.26) \\ \hline
    \end{tabular}
\end{center}
\caption{Table showing $k$, $l$ and $Q$ values for the fits which describe K$_{sim}$/(F$^{2}\alpha^3)$ as a function of the number of sites in the lattice.}
\label{table_Vn3}
\end{table}

\begin{figure}[!htp]
\centering
\includegraphics[width=1\columnwidth]{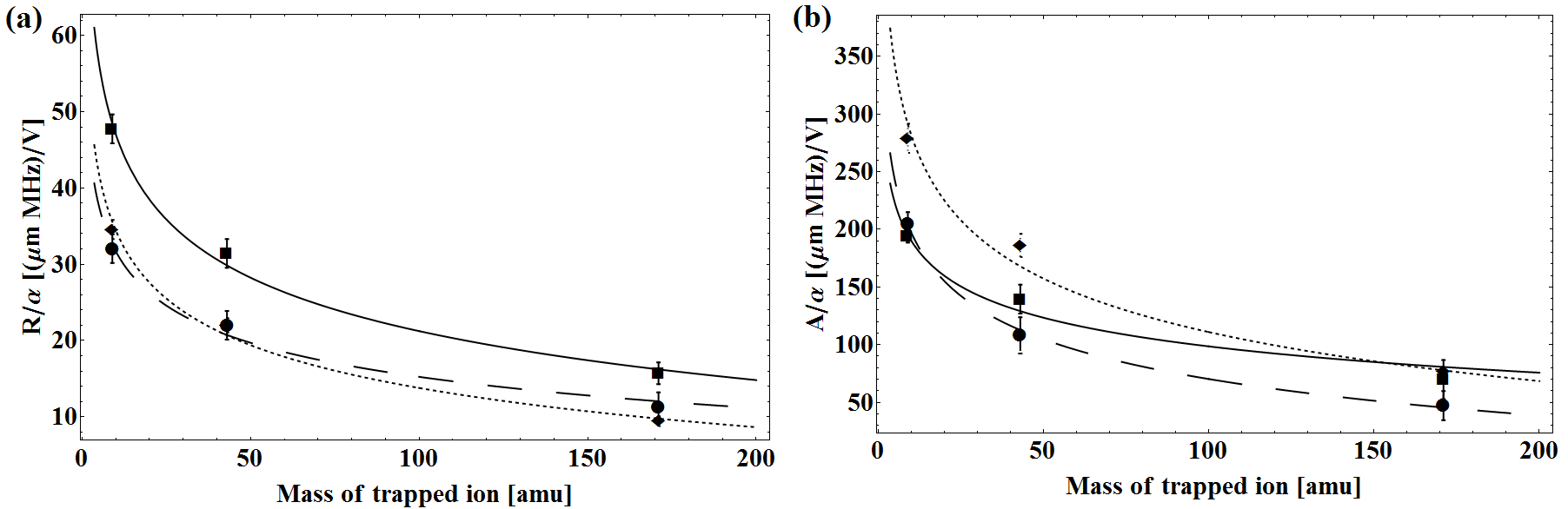}
\caption{(a) Graph showing how the optimum polygon radius varies as a function of the ion mass for 220 (square type unit cells) and 225 (hexagonal and centre rectangular type unit cells) trapping sites. (b)  Graph showing how the optimum polygon separation varies as a function of the ion mass. In both graphs this is shown for square (square markers), hexagonal (circular markers) and centre rectangular (diamond markers) unit cell lattices and the polygon radii and separations are scaled with $\alpha$.}
\label{opto_mass_params}
\end{figure}

\begin{figure}[!htp]
\centering
\includegraphics[width=0.75\columnwidth]{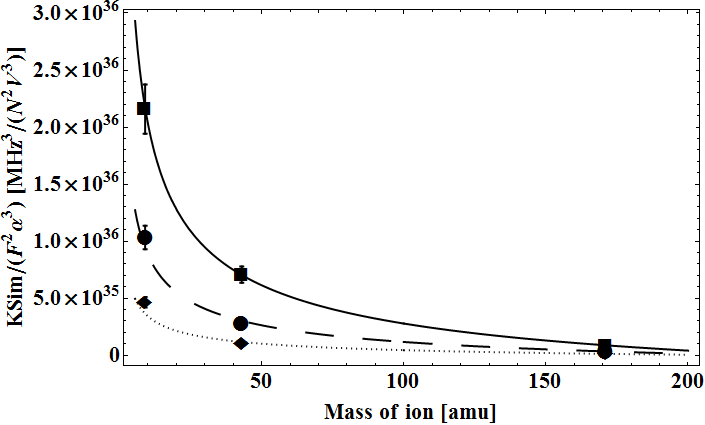}
\caption{Graph showing how the optimum K$_{sim}$/(F$^{2}\alpha^3)$ varies as a function of the ion mass for 220 (square type unit cells (circular markers)) and 225 (hexagonal and centre rectangular type unit cells (square markers and diamond markers respectively)) trapping sites.}
\label{mass_ksim}
\end{figure}

Using the data shown in figures \ref{opto_mass_params}(a) and \ref{opto_mass_params}(b) the optimum radii and separation of the polygons was found to follow a $o+pm^{-0.5}$ and $q+sm^{-0.5}$ relationship, respectively, with values of $o$, $p$, $q$ and $s$ shown in tables \ref{table_Vmass1} and \ref{table_Vmass2}. Figure \ref{mass_ksim} shows how the optimum K$_{sim}$/(F$^{2}\alpha^3)$ varies as a function of the mass of the trapped ion, $m$, for 220 (square type unit cells) and 225 (hexagonal and centre rectangular type unit cells) trapping sites. It is found that the optimum  K$_{sim}$/(F$^{2}\alpha^3)$ scales as $u+vm^{-0.5}$, with the values of $u$ and $v$ shown in table \ref{k_mass_table}.

We note, as shown in figure \ref{opto_mass_params}, that as the mass of the ion is increased, the polygon radii and separation will have to be decreased in order to provide trapping regions with a depth of above 0.1 eV for a given $\alpha$ for 220 (square type unit cells) and 225 (hexagonal and centre rectangular type unit cells) trapping sites. It is also clear to see that ions with lighter masses will provide higher K$_{sim}/(F^2\alpha^3)$ values but will require larger lattice geometries compared to heavier ions.

\begin{table}
\begin{center}
    \begin{tabular}{ | l || c | c | c |}
    \hline
    Lattice Type  & o & p \\ \hline
    Square & -56$\pm$6 & 138$\pm$10 \\ \hline
    Hexagonal & -34$\pm$7 & 88$\pm$11 \\ \hline
    Centre Rectangular & -48$\pm$3 & 110$\pm$6 \\ \hline
    \end{tabular}
\end{center}
\caption{Table showing $o$ and $p$ values for the fits which describe $R/\alpha$ as a function of ion mass.}
\label{table_Vmass1}
\end{table}

\begin{table}
\begin{center}
    \begin{tabular}{ | l || c | c | c |}
    \hline
    Lattice Type  & q & s \\ \hline
    Square & -176$\pm$47 & 489$\pm$65 \\ \hline
    Hexagonal & -311$\pm$15 & 678$\pm$22 \\ \hline
    Centre Rectangular & -402$\pm$28 & 911$\pm$51 \\ \hline
    \end{tabular}
\end{center}
\caption{Table showing $q$ and $s$ values for the fits which describe $A/\alpha$ as a function of ion mass.}
\label{table_Vmass2}
\end{table}

\begin{table}
\begin{center}
    \begin{tabular}{ | l || c | c |}
    \hline
    Lattice Type  & u & v \\ \hline
    Square & -(5.25$\pm$3.00)$\times$10$^{35}$ & (7.96$\pm$1.72)$\times$10$^{36}$ \\ \hline
    Hexagonal & -(1.25$\pm$0.60)$\times$10$^{35}$ & (5.08$\pm$1.29)$\times$10$^{36}$ \\ \hline
    Centre Rectangular & -(2.56$\pm$1.28)$\times$10$^{34}$ & (3.27$\pm$0.86)$\times$10$^{36}$ \\ \hline
    \end{tabular}
\end{center}
\caption{Table showing $u$ and $v$ values for the fits which describe K$_{sim}$/(F$^{2}\alpha^3)$ as a function of the ion mass.}
\label{k_mass_table}
\end{table}

\section{Constraints on $\alpha$}
In this section we will discuss the considerations which could limit the value of $\alpha$. To do this we will show how the power dissipation in a chip trap, the quantum simulation error and the interaction time vary as a function of $\alpha$. This is important as from this a value of $\alpha$ can be determined for a given experiment, which will be shown in section \ref{example_case}.
\subsection{Power dissipation in optimised arrays}
The power dissipation of an ion trap chip is determined by the voltage, $V$, frequency, $\Omega$, as well as the capacitance and resistance of the trap itself. This may, for a given capacitance and resistance, affect the value of $\alpha$ (the ratio between the rf voltage and drive frequency) which can be used. As the value of $\alpha$ is used to determine the optimum polygon radii and separation of a geometry, as shown in figure \ref{opto_params}, for a given number of sites, $N$, and stability parameter, $q$, it is important to know how the power dissipation varies as a function of $\alpha$.

The power dissipation of a chip is approximately given by \cite{Hughes1}

\begin{equation}\label{power_diss}
P_{D}\approx\frac{1}{2}V^{2}C^{2}R\Omega^{2},
\end{equation}

\noindent
where $C$ and $R$ are the capacitance and resistance of the chip. It is not possible to apply any combination of $V$ and $\Omega$ to a geometry as they must be chosen so that the ion is stably trapped with a stability parameter given by \cite{Clark,Madsen}

\begin{equation}\label{q_param_sims2}
q=\frac{2e\eta_{geo} V}{mr^2\Omega^2}
=\frac{2e\eta_{geo}\alpha}{mr^2\Omega}
\end{equation}

\noindent
between zero and 0.9, where $e$ is the charge of an electron.

The ion height, $r$, of ions trapped in the optimised lattices shown in this work have been found to be linearly proportional to $\alpha$. This relationship is shown in figure \ref{thingsValpha} (a) with the constant of proportionality, $k_r$ found to be $\approx$ 60.7 mV$^{-1}$s$^{-1}$ for the example case of square type lattice with 81 sites using $^{171}$Yb$^+$ ions. Considering one particular ion height, $r_0$, and substituting for $r_0=k_r\alpha_0$ and rearranging equation \ref{q_param_sims2} for $\Omega$ yields

\begin{equation}\label{q_param_omega}
\Omega_0=\frac{2e\eta_{geo}}{mk^2_r\alpha_0 q}.
\end{equation}

This equation can be re-expressed for $V_0$ by noting that $V_0=\Omega_0\alpha_0$:

\begin{equation}\label{q_param_volt}
V_0=\frac{2e\eta_{geo}}{mk^{2}_{r}q}.
\end{equation}

Equations \ref{q_param_omega} and \ref{q_param_volt} show that, for a given ion mass, $m$, ion height, $r_0$, stability parameter, $q$, and number of trapping sites in the array (as $k_r$ is a function of the number of trapping sites), there is one unique voltage, $V_0$, and unique parameter $\alpha_0$. This means that a given ion height (and, therefore, a chosen value of $\alpha$) determines both the voltage and drive frequency to be applied to the trap.

To express the power dissipation, $P_D$, in terms of $\alpha$ equation \ref{q_param_omega} and \ref{q_param_volt} can be substituted into equation \ref{q_param_sims2} giving

\begin{equation}\label{alpha_power_diss}
P_D=\frac{8e^4C^2R}{k_r^8m^4q^4\alpha^2}.
\end{equation}

\noindent
Equation \ref{alpha_power_diss} shows that as $\alpha$ is increased, the power dissipated is reduced. This means that the power dissipation is low for high values of $\alpha$ and, as high values of $\alpha$ provide high values of K$_{sim}$ (see figures \ref{opto_ksim} and \ref{mass_ksim}), power dissipation will not impact on producing high values of K$_{sim}$ in optimised geometries. However, a low value of $\alpha$ will result in a high power dissipation in the chip and, so, the maximum allowable power dissipation in a chip will determine the lowest $\alpha$ which can be applied to a geometry.

\subsection{Quantum simulation error}\label{quant_sim_error}
An upper limit on the value of $\alpha$ can be obtained from an estimation of the error of a quantum simulation using the method described in \cite{Deng1}, where the error for the Ising model is given by

\begin{equation}\label{error1}
E_0\approx\frac{1}{2}\eta^2\sum_j(2\overline{n}+1)\left\langle\left[\left[O(t),\sigma_j^z(t)\right],\sigma_j^z(t)\right]\right\rangle.
\end{equation}

\noindent
Here $\overline{n}$ is the mean radial phonon number of the ions, $O$ is the observable of the quantum simulation and $\eta$ is a parameter which characterises phonon displacement caused by the state dependant force and is given by \cite{Deng1}

\begin{equation}\label{eta}
\eta=\frac{F\sqrt{\hbar/(2m\omega)}}{\hbar\omega}
\end{equation}

\noindent
where $m$ is the mass of a trapped ion and $\omega/2\pi$ is its secular frequency.

If $O$ is an $M$-site observable then there exist $M$ non-vanishing commutators (for example a two-site correlation function ($M=2$) or a spin average ($M=1$)) and so the error on the simulation will not be dependant on the number of ions, $N$, in the array \cite{Deng1}. The error in equation \ref{error1} can now be re-written as

\begin{equation}\label{error3}
E_0\approx\frac{F^{2}M(\overline{n}+\frac{1}{2})}{2\hbar m\omega^{3}}.
\end{equation}

Equations \ref{KsimValpha} and \ref{error3} show that both the K$_{sim}$ and the error of the simulation, $E_0$, are proportional to the square of the state dependant force, $F$. It follows that the way in which this force is applied to the ions will determine the dependence of the simulation error on $\alpha$. A discussion on the possible effects of the heating rate on the error can be found in \ref{heating_app}. In this work we will consider applying this state dependant force via laser beams and magnetic field gradients.

To calculate the laser power required to achieve a force, $F$, it is assumed, for illustrative purposes, that the laser beam is focused to a sheet given by 25 $\mu$m multiplied by the width of the array. The laser intensity required to provide a state dependant force, $F$, can be provided by a laser beam of power, $P$. If the output power of the laser used is assumed to be constant, then the force applied to the ions will be dependant on $\alpha$. This is because the lattice size will increase with increasing $\alpha$ and, therefore, so will the spacial area of the beam required to impart a force onto the ions. It is, therefore, required to express this force as a function of $\alpha$. The intensity of a beam required to provide a state dependant force, $F$, is given by \cite{Haljan1}

\begin{equation}\label{intesity_for_F}
I_0=\frac{3F\Delta\lambda I_{sat}}{2\pi\hbar\gamma^2}
\end{equation}

\noindent
where $\Delta$ is the detuning of the laser from resonance, $\lambda$ is the wavelength of the laser, $I_{sat}$ is the saturation intensity of the ion and $\gamma$ is 2$\pi$ times the transition linewidth. The power of a laser beam is given by

\begin{equation}\label{power_beam}
P=I_0a
\end{equation}

\noindent
where $a$ is the spacial area to which the beam is focused. In this work the beam is assumed to be focused to form a light sheet across the array with an area given by $a=(n_s-1)AW=(n_s-1)k_A\alpha W$, where $n_s$ is the number of trapping sites (or polygons) along one side of the array and $W$ is the width of the light sheet. By using equations \ref{intesity_for_F} and \ref{power_beam} the force applied to the ions by a laser power, $P$, can be expressed as

\begin{equation}\label{F_for_power}
F=\frac{2\pi\hbar P\gamma^2}{3a\Delta\lambda I_{sat}}.
\end{equation}

The form of $E_0$ for the case of lasers applying the state dependant force can now be found by using equations \ref{sec_freq_opt}, \ref{F_for_power} and the general error equation \ref{error3} yielding

\begin{equation}\label{E0_laser}
E_{0 laser}=\frac{4\sqrt{2}}{9}\frac{\pi^2\hbar^2}{e^3}\frac{M k_r^6\gamma^4m^2P^2\alpha(\overline{n}+\frac{1}{2})}{\eta_{geo}^3(n_s-1)^2k_A^2W^2\Delta^2\lambda^2I_{sat}^2}.
\end{equation}

\noindent
It follows that both the K$_{sim}$ and simulation error $E_0$ will decrease with increasing laser detuning, $\Delta$. Therefore, the optimum detuning, for a given laser power and $\alpha$, corresponds to the lowest detuning which provides the required K$_{sim}$. The optimum detuning to achieve the lowest simulation error for $^{171}$Yb$^+$ will be discussed in section \ref{sec:Ksim_prime}.

Magnetic fields can also be used to provide the state dependant force, $F$, and is given by \cite{Johanning}

\begin{equation}\label{mag_force}
F_{\hat{i}}=\left(\frac{\hbar}{2}\right)\partial_{i}\omega\left\langle \sigma^{(\hat{i})}\right\rangle
\end{equation}

\noindent
where $\omega=\gamma_{g}bi$ is the position dependant spin resonance frequency with $\gamma_{g}=e/m_{e}$ the gyromagnetic ratio and $i$ is the $x$, $y$ or $z$ direction. The magnetic field gradient $b$ is assumed to arise from a magnetic field of the form $\overline{B}=\overline{B_0}+b\hat{i}$, where $B_0$ is a constant magnetic field offset. From this, the state dependant force, $F_{\hat{i}}$, produced from a magnetic field gradient, $b_{\hat{i}}$, is found to be

\begin{equation}\label{mag_field_grad}
F_{\hat{i}}\approx\frac{\hbar eb_{\hat{i}}}{2m_{e}}
\end{equation}

\noindent
where $m_e$ and $e$ is the mass and charge of an electron respectively. If one assumes the magnetic field is created by a current carrying wire located on the surface of a polygon, at a distance $a$ from the centre of the polygon and making an angle of $45^{\circ}$ with respect to the x-axis then the magnetic field gradient will be of the form

\begin{equation}\label{mag_field_wire}
b_{r^{\prime}}=\frac{\mu_0I}{2\pi r^{\prime 2}}.
\end{equation}

\noindent
Here $\mu_0$ is the permeability of free space, $I$ is the current flowing through the wire and $r^{\prime 2}$ is the distance squared of the ion from the current carrying wire and is equal to $r^2+a^2$ where $r$ is the ion height. We assume, for simplicity, that the distance $a$ scales linearly with $\alpha$ with a constant of proportionality of $k_a$ in order to keep the angle of $r^{\prime}$ to the x-z plane, $\theta$, independent of $\alpha$. As the ion height is known to scale linearly with $\alpha$, from equation \ref{r_alpha}, it is possible to express the magnetic field gradient along $r^{\prime}$ as

\begin{equation}\label{mag_field_wire2}
b_{r^{\prime}}=\frac{\mu_0I}{2\pi\alpha^2\left(k_r^2+k_a^2\right)}.
\end{equation}

\noindent
The component of this magnetic field gradient in the x-z plane can now be shown to be

\begin{equation}\label{mag_field_wire_xzplane}
b_{x,z}=\frac{\mu_0I\cos\theta}{4\pi\alpha^2\left(k_r^2+k_a^2\right)}.
\end{equation}

The form of $K_{sim}$ for the case of magnetic field gradients applying the state dependant force can be found by using equations \ref{sec_freq_opt}, \ref{mag_field_grad}, \ref{mag_field_wire_xzplane} and the general error equation \ref{error3} yielding

\begin{equation}\label{E0_mag}
E_{0 mag}=\frac{\sqrt{2}}{64}\frac{\hbar\mu_0}{\pi^2m_e^2e}\frac{k_r^6m^2I^2\cos^2\theta M\left(\overline{n}+\frac{1}{2}\right)}{\eta_{geo}^3\left(k_r^2+k_a^2\right)^2\alpha}.
\end{equation}

Equations \ref{E0_laser} and \ref{E0_mag} show that the quantum simulation error is proportional to $\alpha$ for a state dependant force created by a laser beam and proportional to $\alpha^{-1}$ for a magnetic field gradient created by current carrying wires. For the case of laser beams the $\alpha$ scaling implies that as $\alpha$ is increased (yielding larger K$_{sim}$ values and geometries as shown in section \ref{sec:optimise_meth}) the quantum simulation error will rise and, therefore, provide an upper limit on the value of $\alpha$.
For the magnetic field gradient case the upper limit on $\alpha$ comes from the current creating the gradient. While the $\alpha$ scaling for the quantum simulation error using magnetic field gradients implies that a larger $\alpha$ is advantageous, the current required to achieve a given magnetic field gradient scales as $\alpha^{2}$ as can be deduced from equation \ref{mag_field_wire2}. The maximum current that one can apply to the lattice therefore provides an upper limit for $\alpha$.

It is also interesting at this point to note the different scaling with $\alpha$ of the laser and magnetic field gradient forces given in equations \ref{F_for_power} and \ref{mag_force}. The laser force can be seen to be $\propto\alpha^{-1}$ as it is a function of the inverse of laser sheet cross section, $a$, which is $\propto\alpha^{1}$. The magnetic gradient force, on the other hand, is $\propto\alpha^{-2}$ as it is a function of the magnetic field gradient $b_{r^{\prime}}$ which is $\propto\alpha^{-2}$ due to $r^{\prime}$ having a linear relationship with $\alpha$ in the geometry considered.
\subsection{Spontaneous emission}\label{sec:Ksim_prime}
When applying the state dependant force to the ions using a laser beam additional decoherence will occur via spontaneous emission. The spontaneous emission rate is given by \cite{Ozeri1, Wineland}

\begin{equation}\label{spont_emission}
S=\frac{\gamma g^2}{6}\left(\frac{1}{\Delta^2}+\frac{2}{\left(\Delta_{fs}-\Delta\right)^2}\right)
\end{equation}

\noindent
where $\gamma$ is 2$\pi$ times the linewidth in Hz, $\Delta$ is the laser detuning from resonance, $\Delta_{fs}$ is the fine structure splitting of the ion ($\approx$ 100 THz for $^{171}$Yb$^+$) and $g$ is the single photon Rabi frequency given by

\begin{equation}\label{photon_rabi}
g=\gamma\sqrt{\frac{I_0}{2I_{sat}}}.
\end{equation}

\noindent
Here, $I_0$ is the laser intensity and $I_{sat}$ is the saturation intensity of the ion. It is possible to express the single photon Rabi frequency in terms of the laser power, $P$, by using equation \ref{power_beam} giving

\begin{equation}\label{photon_rabi_power}
g=\gamma\sqrt{\frac{P}{2(n_s-1)k_A\alpha WI_{sat}}}.
\end{equation}

It is now possible to describe an additional parameter, L$_{sim}$, which describes the ratio of interaction rate to the spontaneous emission rate as

\begin{equation}\label{Lsim_equation}
L_{sim}=\frac{T_S}{T_J}
\end{equation}

\noindent
where

\begin{equation}\label{TNS_equation}
T_{S}=\frac{1}{S}.
\end{equation}

It has been shown that the detuning which minimises the effect of spontaneous emission is $\approx$ 33 THz for $^{171}$Yb$^+$ \cite{Campbell1}. It, therefore, follows that the value of L$_{sim}$ will be maximised with this detuning. This additional parameter is analogous to the parameter K$_{sim}$ in equation \ref{ksim} and is also required to be greater than unity, just like the original K$_{sim}$, when considering a state dependent force created using laser light. If magnetic field gradients are to be used to apply the state dependent force then L$_{sim}$ is no longer relevant.

\subsection{Other considerations}
It is important to note here that an increase in $\alpha$ will increase the time taken for ion-ion interactions to take place in optimised lattice structures, as we will show in equation \ref{interact_time_opt_alpha}. Equation \ref{interact_time1} gives an expression for the time taken for an ion-ion interaction to occur in any given fixed lattice structure. This can be expressed for optimised lattice structures by including the expressions for the ion-ion separation (polygon separation, $A$) and secular frequency, $\omega$, from equations \ref{A_alpha} and \ref{sec_freq_opt} respectively, to yield

\begin{equation}\label{interact_time_opt}
T_J=\frac{e^2\pi\epsilon_0\hbar}{2}\frac{\eta_{geo}^4k_A^3}{k_r^8F^2m^2\alpha}.
\end{equation}

An expression to give the interaction time in optimised lattices as a function of $\alpha$ can now be arrived at by using equations \ref{interact_time_opt} and \ref{F_for_power},

\begin{equation}\label{interact_time_opt_alpha}
T_J=\frac{9}{8}\frac{e^2\epsilon_0}{\pi\hbar}\frac{\eta_{geo}^4k_A^3((n_s-1)k_Aw)^2\Delta^2\lambda^2I_{sat}^2\alpha}{k_r^8\gamma^4P^2m^2}.
\end{equation}

\noindent
Equation \ref{interact_time_opt_alpha} clearly shows that as $\alpha$ is increased the time taken for an ion-ion interaction will increase and, so, it may be preferable to limit the magnitude of $\alpha$ after taking into account the effects on K$_{sim}$. A similar equation can also be derived for the use with magnetic field gradients. It is also important to note here that increasing the laser power, $P$, will increase the value of L$_{sim}$ as the spontaneous emission rate is proportional to $P$ whereas the coupling rate is proportional to $P^2$ as shown in equations \ref{spont_emission} and \ref{interact_time_opt_alpha} respectively. 

With the use of the equations derived in this section, optimal geometries can be calculated given certain experimental parameters and will be described in the following section.
\section{Example case study}\label{example_case}
In this section, an example case will be presented to show how a 2D lattice for the use in quantum simulations can be designed using the work in this paper, whose successful operation is within reach of current technology when using lasers. We then go on to show that while magnetic field gradients can be used to create a state dependant force their use may be more challenging.

From equation \ref{q_param_volt} we can see that there is a unique voltage for a given mass, $m$, lattice type and stability parameter, $q$.
For this example case we choose a lattice comprised of square type unit cells with 9 trapping sites for $^{171}$Yb$^+$ ions with a stability parameter $q=0.5$. Using these parameters the voltage can now be determined by calculating $k_r$. $k_r$ can be found by plotting the ion height of an optimised lattice against $\alpha$, as shown in figure \ref{thingsValpha} (a), and finding the gradient of this linear relationship. For this example case $k_r\approx$ 98 mV$^{-1}$s$^{-1}$. Using this result and equation \ref{q_param_volt} we find the unique voltage to be $\approx$ 34 V where $\eta_{geo}$ has been calculated to be $\approx 0.145$.

The $\alpha$ dependant polygon radius and separation can be calculated by producing the corresponding graphs in figure \ref{opto_params}, using the method described in section \ref{sec:optimise_meth}, for the lattice type and ion to be used. For this example case the optimum radius and separation of the polygons in terms of $\alpha$ is $\approx$ 45$\alpha$ $\mu$m and $\approx$ 174$\alpha$ $\mu$m respectively.

The next step is to choose a laser detuning from resonance and a maximum acceptable error, $E_0$, to be used. For $^{171}$Yb$^+$, as explained in section \ref{sec:Ksim_prime}, the optimum detuning is $\approx$ 33 THz giving a wavelength of $\approx$ 355 nm which is therefore used in this example case together with a maximum acceptable error of 0.25. Note that in this example case the array is assumed to be at cryogenic temperatures which correspond to an electric field noise density three orders of magnitude less than at room temperature. We can see from equation \ref{E0_laser} that $\alpha$ needs to be minimised in order to keep the quantum simulation error low. The minimum $\alpha$ is determined by the lowest ion height one can easily achieve which for this example case we choose to be equal to 30 $\mu m$. For this case we find $\alpha\approx$ 0.3. Having determined $\alpha$ we find the optimum radius and separation to be equal to $\approx$ 14 $\mu m$ and $\approx$ 52 $\mu m$ respectively. To calculate the laser power required equation \ref{E0_laser} should be set to the maximum acceptable error and solved for the laser power, which is found to be $\approx$ 6.5 W assuming the ions to be cooled to $\overline{n}\ll$ 1 and M = 1 (one side average observable). These conditions provide a coupling rate, $J$, of $\approx$ 530 Hz with a $\beta\approx$ 2.8$\times$10$^{-5}$. This laser power can, for example, be achieved using a commercially available diode pumped solid state (Coherent Paladin range) or fibre (Coherent Talisker range) laser system. Table \ref{example_case_table} summarises all parameters for this example case study.

\begin{figure}[!htp]
\centering
\includegraphics[width=0.75\columnwidth]{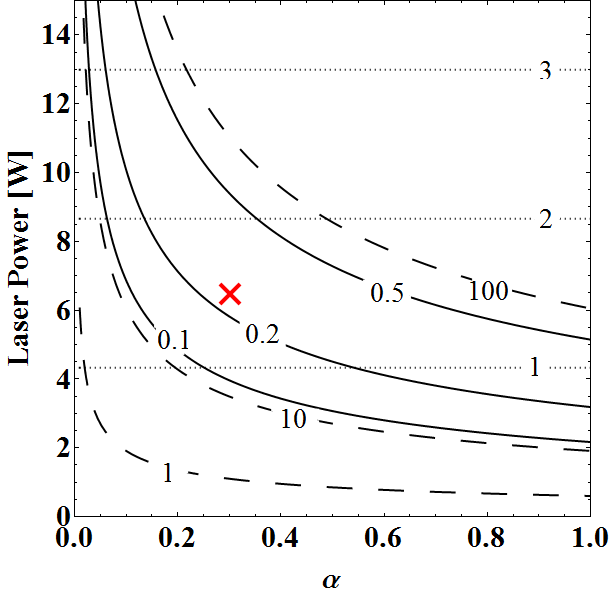}
\caption{Graph showing the quantum simulation error (solid curves), K$_{sim}$ (dashed curves) and L$_{sim}$ (dotted lines) for a three by three square type unit cell lattice with $^{171}$Yb$^+$ ions as a function of $\alpha$ and laser power. Here traps are operated at cryogenic temperature. The cross indicates the example case.}
\label{3x3C}
\end{figure}

Figure \ref{3x3C} shows the effect a change of $\alpha$ and laser power, $P$, has on the quantum simulation error (solid curves), K$_{sim}$ (dashed curves) and L$_{sim}$ (dotted lines). The cross corresponds to the 2D lattice designed in this example case which represents the optimum case in terms of achieving the highest K$_{sim}$ and L$_{sim}$ for a maximum quantum simulation error of 0.25. We note that the main limitations in achieving lower quantum simulation errors stem from the lowest achievable ion height (lowest $\alpha$) and magnitude of electric field noise density. Figure \ref{3x3C} also shows that higher values of K$_{sim}$ and L$_{sim}$ can be achieved with any given geometry (given by the value of $\alpha$) by simply increasing the power of the laser. However, this can only be achieved at the expense of higher quantum simulation errors.

\begin{table}
\begin{center}
    \begin{tabular}{ | c | c | c | c | c | c | c | c | c |}
    \hline
    $\alpha$ & R [$\mu m$] & A [$\mu m$] & $\Delta$ [THz] & P [W] & M & K$_{sim}$ & L$_{sim}$ & $E_{0}$ \\ \hline
    0.3 & 14 & 52 & 33 & 6.5 & 1 & 35 & 1.5 & 0.25  \\ \hline
    \end{tabular}
\end{center}
\caption{Table summarising the parameters for a 3 by 3 square type unit cell lattice at cryogenic temperature as shown in the example case study}
\label{example_case_table}
\end{table}

\begin{figure}[!htp]
\centering
\includegraphics[width=0.75\columnwidth]{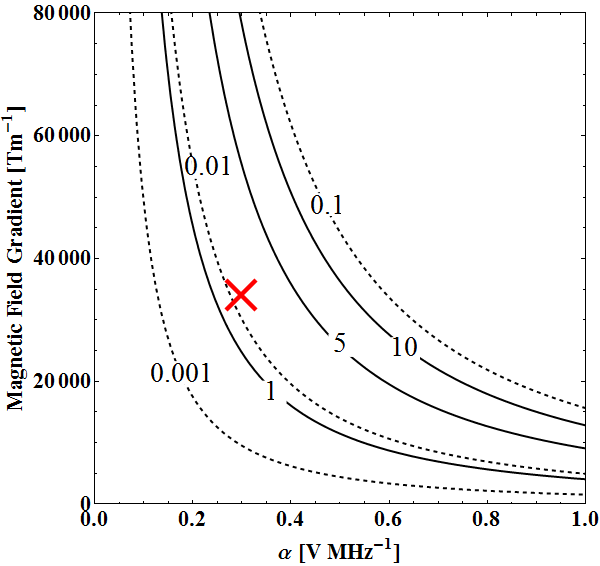}
\caption{Graph showing the quantum simulation error (dashed curves) and the K$_{sim}$ (solid curves) of a three by three square type unit cell lattice with $^{171}$Yb$^+$ ions as a function of $\alpha$ and magnetic field gradient. Here traps are operated at cryogenic temperatures.}
\label{3x3Cmag}
\end{figure}

State dependant forces can also be created using magnetic field gradients as described in section \ref{quant_sim_error}. Figure \ref{3x3Cmag} shows the K$_{sim}$ (solid curves) and the quantum simulation error (dashed curves) as a function of the magnetic field gradient, $b$, and $\alpha$ for traps operated at cryogenic temperature. Here we use $^{171}$Yb$^+$ ions in a three by three square unit cell array. As described in section \ref{quant_sim_error}, $E_0\propto\alpha^{-1}$ indicating that a large $\alpha$ is advantageous. The limit on the maximum $\alpha$ is dependant on the maximum current one can apply to the geometry. Using equation \ref{mag_field_wire2} it is possible to calculate the current required to create a desired magnetic field gradient. In order to illustrate the magnitude of currents required we assume $k_a=k_r$, which will result in an angle $\theta=45^{\circ}$ (refer to section \ref{quant_sim_error} for more information). We also set $\alpha$ to $\approx$ 0.3, determined by the lowest achievable ion height which, for illustration purposes, we have set to 30 $\mu m$. The reason for choosing the minimum $\alpha$ value can be seen when considering equation \ref{mag_field_wire2} which clearly shows that, for a given magnetic field gradient, $I\propto\alpha^2$. In the magnetic field gradient case, the chosen $\alpha$ sets K$_{sim}$ and $E_0$. For this case, again, we choose M = 1, $\overline{n} \ll$ 1, K$_{sim}$ = 2 and $E_0 \approx$ 0.01 which requires a magnetic field gradient of $\approx$ 33,000 Tm$^{-1}$ and is indicated by the cross on figure \ref{3x3Cmag}. This is achievable with a current of $\approx$ 1,200 A yielding a coupling rate, $J$, of $\approx$ 240 Hz and a $\beta\approx$ 2.8$\times$10$^{-8}$. From this simple example case one can conclude that using magnetic field gradients to provide state dependant forces for the use in quantum simulations using the methods and trap designs shown in this work is quite challenging. However, geometries trapping ions in chains allow for sizeable magnetic gradient induced couplings \cite{Welzel1} and, so, a detailed investigation into optimising the wires used for producing magnetic field gradients in the geometries discussed in this work could improve results.

\section{Conclusion}
Two-dimensional arrays of surface ion traps have the potential to provide a technology with which quantum simulations can be performed. In order for ion traps to be used successfully for this purpose a greater understanding is required of how the various geometry parameters affect the ions trapped above them. Throughout this work square, hexagonal and centre rectangular unit cell arrays of microtraps have been modelled in the gap-less plane approximation using the Biot-Savart like law in electrostatics \cite{Mario1}. Decoherence due to motional heating \cite{Turchette} was then compared to the ion-ion interaction \cite{Porras1} to provide a ratio used to describe how much faster an ion-ion interaction occurs in comparison to the motional decoherence, K$_{sim}$. This work investigates how various parameters in the array can be adjusted in order to optimise the device's ability to perform quantum simulations and shows how the interactions can be made as homogeneous as possible over the device. It has been shown how the homogeneity of the K$_{sim}$ across an array can be altered by varying the distance from the outer polygon to the edge of the rf electrode. The distance required to maximise the K$_{sim}$ homogeneity is also shown to vary as a function of the total size of the lattice. The number of polygon sides, $n$, required to maximise K$_{sim}$ has also been found.

We have shown that the K$_{sim}$ of a given lattice geometry can be maximised by reducing the value of $\alpha$. However, as $\alpha$ reduces so does the trap depth. This results in the conclusion that the maximum K$_{sim}$ of a geometry can be achieved by reducing the value of $\alpha$ until the trap depth reaches a reasonable minimum value.

Using this information, optimal geometries as a function of $\alpha$ are presented. This has been achieved by finding the relationships of polygon separation and radius with $\alpha$ for optimal geometries. It was found that, for these optimal geometries, K$_{sim}$ scales as $\alpha^3$. The individual polygon separation and radius were found to posses a linear relationship with $\alpha$ and, therefore, larger geometries have been found to produce larger values of K$_{sim}$. Therefore, the optimal lattice geometry is dependent solely on the value of $\alpha$ for a given ion mass and number of trapping sites in the array.

We presented a case study for determining an optimum geometry consisting of 9 trapping sites arranged into square type unit cells for $^{171}$Yb$^+$ ions. We showed how the value of $\alpha$ can be chosen (and, therefore, the geometry dimensions) by taking into account the laser power (or static magnetic field gradient) required to produce a state dependant force acting on the ions, the K$_{sim}$ and the error on the simulation. From this it has been found that to carry out quantum simulations with reasonable K$_{sim}$ and error values it is preferable to use traps held at cryogenic temperatures as this reduces decoherence due to heating effects on the ions. Other methods known to significantly decrease the anomalous heating rate include pulsed laser electrode cleaning \cite{Allcock2} and Argon-ion beam electrode cleaning \cite{Hite}.

The scaling of anomalous heating has not yet been fully understood and is thought to posses a dependence with the ion height, $r$, of between $r^{-4}$ and $r^{-2}$. In this work we have used $r^{-4}$ however, if the anomalous heating is found to posses a relationship with ion height nearer $r^{-2}$ then the equations in this work can be adjusted and this is discussed in more detail in \ref{Appendix}.

The relationships between lattice size and $\alpha$ with the polygon radii and separation obtained using the method described in this work will allow for the construction of two-dimensional surface trap lattice arrays with high ratios of ion-ion interaction rates to decoherence rates, providing a system which could be used to perform quantum simulations.

\section*{Acknowledgements}
We are pleased to acknowledge useful discussions with Simon Webster. This work is supported by the UK Engineering and Physical Sciences Research Council (EP/E011136/1, EP/G007276/1), European Community's Seventh Framework Programme (FP7/2007-2013) under grant agreement no. 270843 (iQIT), the European Commission's Sixth Framework Marie Curie International Reintegration Programme (MIRG-CT-2007-046432), the Nuffield Foundation and the University of Sussex.
\appendix
\section{}\label{heating_app}
The simulation error in equation \ref{error3} can be adjusted to take into account the heating of the ions during a simulation. If this occurs the mean radial phonon number $\overline{n}$ will become time dependent, $\overline{n}(t)$. It, therefore, follows that the error will also become time dependent and is given by

\begin{equation}\label{appe1}
E_0(t)\approx\frac{F^{2}M(\overline{n}(t)+\frac{1}{2})}{2\hbar m\omega^{3}}.
\end{equation}

The time dependent mean radial phonon number will be a function of the heating rate $\dot{n}$, the interaction time, $T_J$, and the initial mean phonon number, $\overline{n}_0$ and is given by

\begin{equation}\label{appe2}
\overline{n}(T_J)=\overline{n}_0+\dot{n}T_J
\end{equation}

\noindent
yielding an error given by

\begin{equation}\label{appe3}
E_0(T_J)\approx\frac{F^{2}M(\overline{n}_0+\dot{n}T_J+\frac{1}{2})}{2\hbar m\omega^{3}}.
\end{equation}

\noindent
It should be noted here that $\dot{n}T_J=1/K_{sim}$ and so the error can be expressed as

\begin{equation}\label{appe4}
E_0(T_J)\approx\frac{F^{2}M(\overline{n}_0+\frac{1}{K_{sim}}+\frac{1}{2})}{2\hbar m\omega^{3}}.
\end{equation}

Equation \ref{appe4} shows that as $K_{sim}$ increases the error will tend towards that in equation \ref{error3} in section \ref{quant_sim_error} for the case of $\bar{n}\ll 1$. This is because higher values of $K_{sim}$  mean that less heating takes place during an interaction until the mean radial phonon number can be approximated as constant. 

\section{}\label{Appendix}
In the work presented we have used a motional heating rate, $\dot{n}$, which has an $r^{-4}$ scaling, where $r$ is the ion height. However the scaling of this anomalous heating has not yet been fully understood and so it is conceivable that a different scaling will be required to describe the effect. For this reason it is the aim of this appendix to outline the steps and main expressions required to obtain an optimised two-dimensional ion trap array with an anomalous heating rate which posses an arbitrary scaling with the ion height $r^{-x}$.

For the case of optimising the homogeneity of $K_{sim}$ across an array the results shown in section \ref{sec:optimise_g} will hold for any scaling of the heating rate with ion height. This is because this optimisation aims to give each trapping site in the array the same properties and this is independent of the heating rate. The same applies to the optimisation of the number of polygon sides described in section \ref{sec:optimise_n}.

Equation \ref{KsimValpha} describing $K_{sim}$ for an optimised geometry can be altered to take into account a different scaling of the heating rate with ion height. For an arbitrary scaling $r^{-x}$ this equation can be expressed as

\begin{equation}\label{ksim_app}
K_{sim}=\frac{4F^2mk_r^{\left(4+x\right)}\alpha^{\left(x-1\right)}}{\Xi\eta^2_{geo}\pi\epsilon_{0}k_A^3}.
\end{equation}

It can be seen in equation \ref{ksim_app} that the $K_{sim}$ is proportional to $\alpha^{\left(x-1\right)}$. The values of $k_r$, $k_A$ and $k_R$ (described in section \ref{sec:optimise_meth}) are independent of value of $x$ and, so, the optimum geometry for a given $\alpha$ will be the same regardless of the scaling of the heating rate with ion height. Using this procedure optimum geometries can be computed for an arbitrary scaling of the anomalous heating with the ion height, $r^{-x}$.

\section*{References}

\end{document}